\documentclass[aps,prl,twocolumn,amsmath,showpacs]{revtex4}
\usepackage{graphicx}
\begin{document}
\title{Three-boson problem near a narrow Feshbach resonance} 
\author{D. S. Petrov}
\affiliation{ITAMP, Harvard-Smithsonian Center for Astrophysics, Cambridge,
Massachusetts 02138, USA\\
Harvard-MIT CUA, Physics Department, Harvard University, Cambridge, Massachusetts
02138, USA\\
Russian Research Center, Kurchatov Institute, Kurchatov Square, 123182 Moscow,
Russia} 
\date{\today}
\begin{abstract}

We consider a three-boson system with resonant binary interactions and show that for
sufficiently narrow resonances three-body observables depend only on the resonance
width and the scattering length. The effect of narrow resonances is qualitatively
different from that of wide resonances revealing novel physics of three-body
collisions. We calculate the rate of three-body recombination to a weakly bound
level and the atom-dimer scattering length and discuss implications for experiments
on Bose-Einstein condensates and atom-molecule mixtures near Feshbach resonances.

\end{abstract}

%
\maketitle

The creation of strongly interacting atomic BEC's by increasing the scattering
length, $a$, is impeded by inelastic processes \cite{Ketterle1999}, the most
important of which is three-body recombination to molecular states (see
\cite{Fedichev3body1996}, and references therein). For several years this process has
been the primary motivation to study three-body physics in the context of ultracold
gases. Recent observations of weakly bound dimers and their mixtures with atoms
\cite{GrimmMolecules,RempeMolecules,KetterleMolecules} stimulate a strong interest
in atom-dimer collisions and in peculiar weakly bound trimer states (Efimov states)
\cite{Efimov}, which have been extensively studied in theory but have never been
observed experimentally.

In the case of a potential resonance when the scattering length is much larger than
the radius, $R_e$, of interatomic forces, low-energy observables in a three-boson
system are universal functions of $a$ and the three-body parameter $r_0$. In
particular, for $a>0$ the rate constant of three-body recombination of atoms of mass
$m$ to a weakly bound state is given by
$\alpha_{\rm rec}=C\hbar a^4/m$, where $C$ is a periodic function of $\log(a/r_0)$
and
can have any value between $0$ and approximately $68$ \cite{Recomb,BraatenRecomb}.
The atom-dimer scattering length is periodic in $\log(a/r_0)$ and goes through
infinity each time a new Efimov state is at the atom-dimer threshold
\cite{BraatenAtomDimer,EfimovAtomDimer}. The three-body parameter
$r_0$ may be determined from the three-body wavefunction at
distances of the order of $R_e$. For realistic interatomic potentials an
{\it ab initio} calculation of $r_0$ is difficult and it is usually considered as
a parameter of the theory. The related uncertainty makes an exact determination of
the three-body observables impossible. Moreover, application of these results to
three-body systems near a narrow Feshbach resonance should be done with care as the
two-body scattering then depends strongly on energy. 

Feshbach resonances occur when the energy, $E$, of a pair of colliding atoms in the
open channel is close to the energy, $E_{res}$, of a quasi-discrete
molecular state in the closed channel \cite{VerhaarFeshbach}. In the limit when both
$E$ and $E_{res}$ are much smaller than the spacing, $D\sim \hbar^2/mR_e^2$, between
molecular levels, and the detuning $E-E_{res}$ is sufficiently small to neglect the
background scattering \cite{footnote1}, the scattering amplitude is given by
\cite{LLQ}:
\begin{equation}\label{twobody.conv}
F(E)\approx-\frac{\hbar\gamma/\sqrt{m}}{E-E_{res}+i\gamma\sqrt{E}}=
-\frac{1}{1/a+R^*q^2+iq},
\end{equation}
where $q=\sqrt{mE}/\hbar$, $a=-\hbar\gamma/\sqrt{m}E_{res}$, and we introduce the
length $R^*$ related to the width of the resonance by $R^*=\hbar/\sqrt{m}\gamma>0$
\cite{footnote2}. 

In this Letter we show that the resonant dependence of the scattering amplitude on
$E$ is of fundamental importance for the three-body problem and develop a method for
its solution near a narrow resonance in the case $R^*\gg R_e$, $|a|\gg R_e$. The
three-body recombination rate, the positions of Efimov states, and the
atom-dimer scattering length are shown to be functions only of the two-body
observables, $a$ and $R^*$. We identify two physically distinct regimes of the
three-body scattering. In the regime of small detuning, $|a|\gg R^*$, the physics of
a three-body collision is the same as near a wide resonance. However, $r_0$ is
unambiguously determined by the parameter $R^*$. The regime of intermediate
detuning, $|a|\ll R^*$, is qualitatively different. There are no Efimov states for
three bosons and for $a>0$ neither the atom-dimer scattering length nor the
three-body recombination rate shows the periodic dependence on $\log(a)$. The rate
constant of three-body recombination to a weakly bound level is proportional to
$a^{7/2}R^{*1/2}$ and by far exceeds the maximum value expected from the $a^4$-law
\cite{Recomb,BraatenRecomb}. Our result for $\alpha_{\rm rec}$ agrees with the
experimental observation at MIT \cite{Ketterle1999}.

First, we analyze the consequences of the finite resonance width for binary
collisions. Although the validity of Eq.~(\ref{twobody.conv}) does not require
$qR^*\ll 1$, the amplitude (\ref{twobody.conv}) formally corresponds to a
truncated effective range expansion with the range $R=-2R^*$. In the case $R^*\gg
R_e$ this range exceeds the actual radius of interatomic forces and the interaction
acquires unusual {\it long-range} properties. Two examples of such resonances are
the 907 G Feshbach resonance in Na ($\Delta_B\approx 1$ G, $R_e\approx 45$\AA\,,
$R^*\approx 260$\AA\,) \cite{Ketterle1999} and the 1007.4 G resonance in $^{87}$Rb
($\Delta_B\approx 0.17$ G, $R_e\approx 85$\AA\,, $R^*\approx 320$\AA\,)
\cite{RempeFeshbach,RemarkOther}. The amplitude (\ref{twobody.conv}) is also a
feature of optically induced resonances. A pair of free atoms can be coupled to an
electronically excited molecular level \cite{FedichevResLight} or, through a
two-photon Raman transition reducing the spontaneous emission, to a vibrational
state of the ground-state potential \cite{RemarkTwoPhoton}. In contrast to magnetic
resonances, the width of an optical one can be modified by changing laser
intensities. 

The poles of the scattering amplitude (\ref{twobody.conv}) in the
complex energy plane determine the positions of weakly bound molecular states. From
the quadratic equation, the inverse size of the bound state is
\begin{equation}\label{twobody.boundstateenergy}
\kappa=\sqrt{m |E_b|}/\hbar=(\sqrt{1+4R^*/a}-1)/2R^*.
\end{equation}
A true bound level exists only for $a>0$. In the regime of small detuning we
retrieve the well-known result $\kappa\approx 1/a$. In the regime of intermediate
detuning, $\kappa\approx 1/\sqrt{aR^*}$ and the binding energy is close to
$E_{res}=-\hbar^2/maR^*$. Remarkably, due to the inequality $R^*\gg R_e$
this level remains weakly bound even for $a\sim R_e$. 

Outside the radius of the potential the normalized wavefunction of the bound state is
given by
\begin{equation}\label{twobody.boundstate}
\phi_b(r)=(1+2\kappa R^*)^{-1/2}\sqrt{\kappa/2\pi}\,\exp(-\kappa r)/r,
\end{equation}
where the preexponential factor is determined by the residue of the scattering
amplitude at $E=E_b$ \cite{LLQ}. The normalization integral calculated with the
wavefunction (\ref{twobody.boundstate}) tends to $1$ in the regime of small
detuning. We conclude that atoms in the dimer are well separated and spend most of
their time in the open channel. As the bound state becomes deeper the normalization
integral decreases that reflects increasing occupation of the quasi-stationary level.
In the regime of intermediate detuning the weakly bound state is this molecular
level with a small admixture of the open-channel wavefunction. 

To solve the three-body problem we employ the zero-range approximation valid if the
binary collision energies are small compared to $D$. The key idea is to solve the
equation for free motion but with the Bethe-Peierls boundary condition on the
wavefunction for vanishing distance, $r$, between two atoms:
$-(r\psi)'/r\psi\rightarrow \tilde{a}^{-1}(E_{c})$, where the energy-dependent
scattering length defined as $\tilde{a}^{-1}(E)=a^{-1}+R^*mE/\hbar^2$ should be
evaluated at the collision energy, $E_{c}$, of these two atoms. This boundary
condition, which can be also rewritten as $\psi\propto 1/r-1/\tilde{a}(E_{c})$,
leads to the scattering amplitude (\ref{twobody.conv}).

In the center of mass reference frame three bosons with total energy $E$ are
described by the equation
\begin{equation}\label{general.poisson}
-[\nabla_{\bf x}^2+\nabla_{\bf y}^2+E]\Psi({\bf x},{\bf y})=0
\end{equation}
where ${\bf y}$ is the distance between two bosons, $\sqrt{3}{\bf x}/2$ is the
distance between their center of mass and the third atom, and $m=\hbar=1$. Due
to the bosonic symmetry $\Psi$ is invariant under the transformations:
\{${\bf x}\rightarrow (\pm\sqrt{3}\,{\bf y}-{\bf x})/2$, ${\bf y}\rightarrow
(\sqrt{3}\,{\bf x}\pm{\bf y})/2$\}, and \{${\bf x}\rightarrow {\bf x}$, ${\bf
y}\rightarrow -{\bf y}$\}. Therefore, three boundary conditions for $y\rightarrow 0$
and for $\sqrt{3}\,{\bf x}\pm{\bf y}\rightarrow 0$ are dependent and we need
to ensure the proper behavior of $\Psi$ only for $y\rightarrow 0$:
\begin{equation}\label{twobody.boundary}
\Psi({\bf x},{\bf y})\approx [y^{-1}-\tilde{a}^{-1}(E_{c})]f({\bf x})/4\pi.
\end{equation}
The function $f$ contains the information about the relative motion of the third atom
with respect to the first two when they are on top of each other. The quantity 
$E_{c}$ is the total energy $E$ excluding the energy of this relative motion.
Accordingly, $E_{c}$ can be presented as a differential operator acting on $f$:
\begin{equation}\label{general.q}
E_{c}=\lim_{{\bf y}\rightarrow 0}\frac{-\nabla_{\bf y}^2\Psi}{\Psi}=E+\lim_{{\bf
y}\rightarrow 0}\frac{\nabla_{\bf x}^2\Psi}{\Psi} =E+\frac{\nabla_{\bf x}^2f}{f}.
\end{equation} 
Since $\tilde{a}^{-1}(E_{c})=a^{-1}+R^*E_{c}$, Eq.~(\ref{twobody.boundary}) reduces
to
\begin{equation}\label{general.boundary}
\Psi\approx \left[y^{-1}-a^{-1}-R^* (E+\nabla_{\bf x}^2)\right]f({\bf x})/4\pi.
\end{equation}

Let us now consider the following solution of Eq.~(\ref{general.poisson}):
\begin{align}\label{general.psi}
\Psi({\bf x},{\bf y}) &=\Psi_0+\int{\rm d}^3r'f({\bf r'})\bigl[G_E(\!\sqrt{({\bf
x}-{\bf r'})^2+y^2}) \nonumber \\
 + & \sum_\pm G_E\Bigl(\!\sqrt{({\bf x}-{\bf r'}/2)^2+({\bf
y}\pm\sqrt{3}{\bf r'}/2)^2}\Bigr)\bigr],
\end{align} 
where $\Psi_0({\bf x},{\bf y})$ is a properly symmetrized and finite solution of
Eq.~(\ref{general.poisson}) without singularities and $G_E$ is the Green function of
Eq.~(\ref{general.poisson}) given by $G_E(X)=-EK_2(\sqrt{-E}X)/8\pi^3\!
X^2$. For negative energies $K_2$ is an exponentially decaying Bessel function and
for $E>0$ we use the convention $\sqrt{-E}=-i\sqrt{E}$. A direct examination shows
that the wavefunction (\ref{general.psi}) is properly symmetrized. In order
to ensure that (\ref{general.psi}) reduces to (\ref{general.boundary}) at
$y\rightarrow 0$ we add and subtract from Eq.~(\ref{general.psi}) an auxiliary
quantity
$$
f({\bf x})\!\int\! G_E(\!\sqrt{({\bf x}-{\bf r'})^2+y^2}){\rm d}^3r'=f({\bf
x})e^{-\sqrt{-E}y}/4\pi y.
$$
As $y$ tends to zero we see that the terms proportional to $1/y$ in
Eqs.~(\ref{general.boundary}) and (\ref{general.psi}) are equal and matching the next
(regular) terms yields the equation for the function $f$:
\begin{equation}\label{general.main}
(-R^*\nabla_{\bf r}^2+\hat L_E-a^{-1}+\sqrt{-E}-R^*E)f({\bf r})=F_0({\bf r}),
\end{equation}
where $F_0({\bf r})=4\pi \Psi_0({\bf r},0)$ and $\hat L_E$ is given by
\begin{eqnarray}\label{general.L}
\hat L_E f({\bf r})&=&4\pi\int\big\{G_E(|{\bf r}-{\bf r'}|)[f({\bf r})-f({\bf
r'})]\nonumber \\
&-&2\, G_E(\!\sqrt{r^2+r'^2+{\bf
r}{\bf r'}})f({\bf r'})\big\}{\rm d}^3r'.
\end{eqnarray} 
The operator on the left hand side of Eq.~(\ref{general.main}) conserves angular
momentum, and we can expand $f$ in spherical harmonics to work only with a set of
uncoupled equations for functions of a single variable $r$. In momentum space $f({\bf
k})=\int f({\bf r})\exp(i{\bf kr}){\rm d}^3r$ and $\hat L_E$ is given by
\begin{eqnarray}\label{general.Lmomentum}
\hat L_E f({\bf k})&=&(\sqrt{-E+k^2}-\sqrt{-E})f({\bf k})\nonumber \\
&-&\frac{2}{\sqrt{3}\pi^2} \int \frac{f({\bf q})\,{\rm
d}^3q}{k^2+q^2+{\bf k}{\bf q}-3E/4}.
\end{eqnarray}
 
Equation ({\ref{general.main}) was first obtained in Ref.~\cite{Skorniakov} for
$R^*=0$. In that case its solution is not unique \cite{DanilovAtomDimer} -- at small
distances it behaves as $f\sim r^{-1}\sin\log(r/r_0)$ with arbitrary $r_0$.
According to Eq.~(\ref{general.q}) the corresponding collision energy increases as
$|E_c|\sim 1/r^2$ at small $r$, and for narrow resonances the two-body collisions
would become off-resonant at $r\sim R^*$. We show below that the solution in this
case is smooth and unique. Equations ({\ref{general.main}) and (\ref{general.psi})
determine the three-body wavefunction and all low-energy properties of the system. 

Consider the problem of an atom scattering by a weakly bound dimer ($a>0$) at
low collision energies when the scattering amplitude is dominated by the $s$-wave
contribution. The corresponding scattering length, $a_{ad}$, is determined by the
long-range asymptote of the atom-dimer wavefunction at zero collision energy.
Namely, in the region $y\lesssim 1/\kappa\ll x$ the total three-body wavefunction
factorizes into $\Psi({\bf x},{\bf y})\approx \phi_b(y)(1-2a_{ad}/\sqrt{3}x)$.
A comparison of this expression with Eqs.~(\ref{twobody.boundstate}) and
(\ref{twobody.boundary}) gives $f(r)\propto 1-2a_{ad}/\sqrt{3}r$ at large $r$.
Thus, we can find $a_{ad}$ by solving Eq.~({\ref{general.main}) with $E=E_b$:
\begin{equation}\label{atomdimer.main}
(-R^*\nabla_{\bf r}^2+\hat L_{E_b})f(r)=0.
\end{equation}
To obtain Eq.~({\ref{atomdimer.main}) from Eq.~({\ref{general.main}) we note
that $-a^{-1}+\sqrt{-E_b}-R^*E_b=0$ and $\Psi_0\equiv 0$ for $E<0$.

At distances $r\ll 1/\kappa$ we can approximate the Green function in
Eq.~(\ref{general.L}) by $G_{E_b}(X)\approx 1/4\pi^3 X^4=G_0(X)$.
This gives $\hat L_{E_b}\approx \hat L_0$. The operator $\hat L_0$ has the property
that 
$\hat{L}_0r^\nu=\lambda(\nu)r^{\nu-1}$ for $-3<{\rm Re}\,(\nu)<1$. The function
$\lambda(\nu)$ has two roots, $\nu_\pm=-1\pm is_0$, where $s_0\approx 1.00624$.
Thus, the solution of Eq.~(\ref{atomdimer.main}) for $R^*=0$ and $r\ll
1/\kappa$ is 
\begin{equation}\label{atomdimer.asym}
f(r)\propto r^{-1}\sin[s_0\ln(r/r_0)]. 
\end{equation}

In the regime of small detuning Eq.~(\ref{atomdimer.asym}) remains valid at
distances $R^*\ll r\ll a$, where the Laplacian in Eq.~(\ref{atomdimer.main})
can be neglected. However, for $r\ll R^*$ the differential operator is dominant and
$f$ should tend to a finite value \cite{RemarkFiniteValue}. Then, the integral
operator acts as a Coulomb potential $\hat L_0\approx \lambda(0)/r$, where
$\lambda(0)=-4/\sqrt{3}$. Note that the two-body collision energy $|E_c|\lesssim
1/R^*R_e\ll D$, which justifies the use of the zero-range method. 

Expression (\ref{atomdimer.asym}) and scaling properties of
Eq.~(\ref{atomdimer.main}) at small distances show that the long-range asymptote
of $f$ does not change if $R^*$ is multiplied or divided by $\exp(\pi/s_0)\approx
22.7$, provided that $R^*$ remains much smaller than $a$. Obviously, the atom-dimer
scattering length has the same property. Numerical integration of
Eq.~(\ref{atomdimer.main}) in the limit $R^*\ll a$ yields $r_0\approx 0.5 R^*$.

In the regime of intermediate detuning the integral operator $\hat L_{E_b}$ in
Eq.~(\ref{atomdimer.main}) is a perturbation to the differential one with
the small parameter $1/(R^*\kappa)\approx\sqrt{a/R^*}$. In the zeroth
order $f=const$, i.e. there is no scattering. The first two corrections yield
\begin{equation}\label{atomdimer.Rgga}
a_{ad}/a\approx -8/3[1+(11/6)\sqrt{a/R^*}].
\end{equation}

In Fig.~\ref{amol} we plot $a_{ad}/a$ as a function of $R^*/a$ calculated numerically
from Eq.~(\ref{atomdimer.main}). The resonances in $a_{ad}$ occur when there is a
Efimov state of zero energy relative to the atom-molecule continuum. We find that the
first Efimov state arises at $R^*/a\approx 2.2$, the second -- at $R^*/a\approx
0.07$. With high accuracy $a_{ad}/a$ is periodic on a logarithmic scale as we
further decrease $R^*/a$. At large $R^*/a$ we recover the analytic result
(\ref{atomdimer.Rgga}).

\begin{figure}[h]
\includegraphics[width=\hsize]{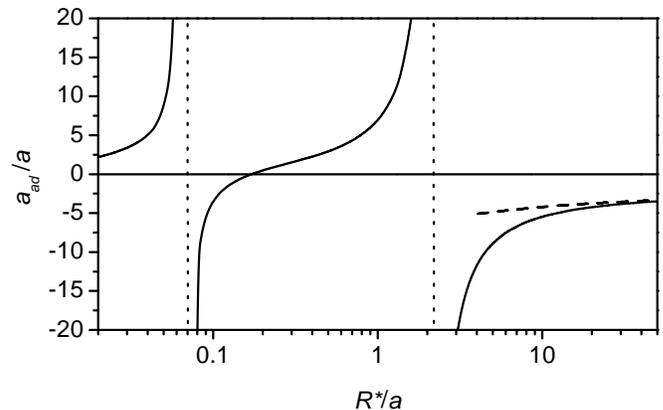}
\caption{
 \label{amol}
The ratio $a_{ad}/a$ versus $R^*/a$.  Dashed curve is given by
Eq.~(\ref{atomdimer.Rgga}) and corresponds to the limit $a\ll R^*$. Dotted lines
indicate the points $R^*/a\approx 2.2$ and $R^*/a\approx 0.07$, where the first and
the second Efimov states emerge.}
\end{figure}

We now turn to the problem of three-body recombination to a weakly bound level. This
process occurs when three atoms approach each other to distances of the order of
$1/\kappa$. Therefore, at energies $E\ll |E_b|$ the recombination probability is
energy
independent and we can consider the zero-energy limit of Eq.~(\ref{general.main}).
Let the function $\Psi_0$ describe three noncondensed ideal bosons in a unit volume.
Then in the region important for the recombination $\Psi_0\approx
1/\sqrt{6}$ and Eq.~(\ref{general.main}) acquires the form
\begin{equation}\label{rec.main}
(-R^*\nabla_{\bf r}^2+\hat L_0-a^{-1})f(r)=4\pi/\sqrt{6}.
\end{equation}
The solution of Eq.~(\ref{rec.main}) at large $r$ contains a term $A\exp(i\kappa
r)/r$. By virtue of Eqs.~(\ref{twobody.boundstate}) and (\ref{twobody.boundary}) the
corresponding part of the three-body wavefunction is $\Psi({\bf x},{\bf y})\approx
A\sqrt{(1+2\kappa R^*)/8\pi \kappa}\,\phi_b(y)\exp(i\kappa x)/x$ and describes an
atom and a dimer flying apart. Given the factor $A$, the derivation of the rate
constant of three-body recombination $\alpha_{\rm rec}$ is straightforward. We solve
Eq.~(\ref{rec.main}) in momentum representation and determine $A$ by the residue of
the function $f(k)$ at the pole $k=\kappa$. 

In Fig.~\ref{3body706} we plot $\alpha_{\rm rec}/a^4$ as a function of $R^*/a$
calculated
numerically from Eq.~(\ref{rec.main}). In the regime of small detuning our approach
leads to the periodic dependence of $\alpha_{\rm rec}/a^4$ on $\log(a)$ with the
maximum
value $\approx 68$ consistent with Refs.~\cite{Recomb,BraatenRecomb}. In fact, our
results for $a_{ad}$ and for $\alpha_{\rm rec}$ in this limit coincide with those
obtained by using the effective field theory \cite{BraatenAtomDimer,BraatenRecomb}.
The three-body parameter, $\Lambda_*$, introduced there is related to $R^*$ by
$\Lambda_*\approx 6.6/R^*$. 

In the regime of intermediate detuning the operator $R^*\nabla_{\bf r}^2$ in
Eq.~(\ref{rec.main}) dominates and we calculate $\alpha_{\rm rec}$ perturbatively.
Expanding in $\sqrt{a/R^*}$, the first two terms are
\begin{equation}\label{rec.Rgga}
\alpha_{\rm rec}\approx 192\sqrt{3}\pi^2
\sqrt{a^7R^*}\left[1-\left(\frac{4\pi}{3\sqrt{3}}-1\right)\sqrt{\frac{a}{R^*}}\right].
\end{equation}
Our results agree with measurements at MIT \cite{Ketterle1999}. In particular,
$\alpha_{\rm rec}/a^4\gg 68$ in this regime.

\begin{figure}[h]
\includegraphics[width=\hsize]{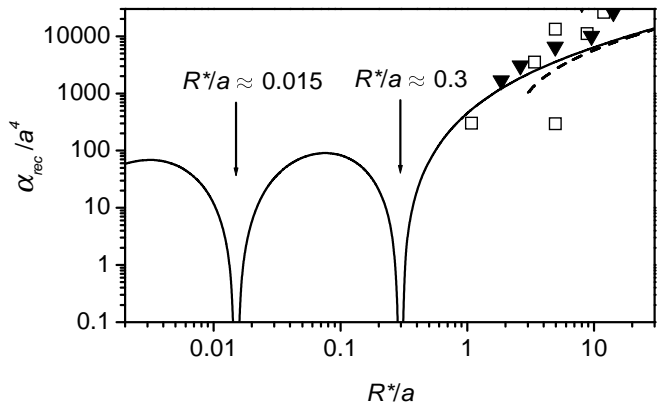}
\caption{
 \label{3body706}
The ratio $\alpha_{\rm rec}/a^4$ versus $R^*/a$.  Dashed curve is given by
Eq.~(\ref{rec.Rgga}). Arrows indicate values of $R^*/a$ at which $\alpha_{\rm rec}$
vanishes. Scatter points correspond to experimental data taken near the 907 G
Feshbach resonance in Na at MIT \cite{Ketterle1999}.}
\end{figure}

Finally, we point out that on the magnetic field axis the region of intermediate
detuning is a factor of $R^*/R_e$ wider than the region of small detuning and is,
therefore, more accessible. Note, that in both MIT \cite{Ketterle1999} and MPIQ
\cite{RempeFeshbach} experiments on narrow resonances $a\lesssim R^*$. 

The method presented here can be generalized to describe fermions and fermion-boson
mixtures with different masses. The significant dependence of two- and three-body
observables on $R^*$ suggests that properties of a many-body system should also
depend on $R^*$.

The atom-dimer scattering can be probed by a direct collision of
atomic and molecular clouds. It is remarkable that near the wide resonance associated
with the first Efimov state ($a\approx 0.45R^*$) the atom-dimer scattering length
can be tuned to be an order of magnitude larger than $a$. In a degenerate
atom-molecule mixture one has either a phase separation or a collapse depending on
the sign of $a_{ad}$. We may also consider the formation of trimers by adiabatic
sweep across this resonance. Their detection can be based on the Stern-Gerlach
technique \cite{GrimmMolecules,RempeMolecules,KetterleMolecules} as atoms, dimers
and trimers all have different magnetic moments. The vanishing of the
recombination to the weakly bound level at $a\approx 3.3R^*$ can be used to produce
a long-lived strongly interacting atomic gas.

We thank V. Kharchenko, A. Dalgarno, M. Lukin, and R. Krems for fruitful discussions,
J. Stenger for sending the data
of Ref.~\cite{Ketterle1999} in numerical form, and E. van Kempen for providing the
parameters of the 1007 G resonance in Rb. This work was supported by NSF through a
grant for the Institute for Theoretical Atomic, Molecular and Optical Physics at
Harvard University and Smithsonian Astrophysical Observatory and
by the Russian Foundation for Basic Studies.


\begin{references} 

\bibitem{Ketterle1999} J. Stenger, S. Inouye, M.R. Andrews, H.-J. Miesner, D.M.
Stamper-Kurn, and W. Ketterle, Phys. Rev. Lett. {\bf 82}, 2422 (1999).

\bibitem{Fedichev3body1996} P.O. Fedichev, M.W. Reynolds, and G.V. Shlyapnikov, Phys.
Rev. Lett. {\bf 77}, 2921 (1996).

\bibitem{GrimmMolecules} J. Herbig, T. Kraemer, M. Mark, T. Weber, C. Chin, H.C.
Nagerl, and R. Grimm, Science {\bf 301}, 1510 (2003).

\bibitem{RempeMolecules} S. D\"urr, T. Volz, A. Marte, and G. Rempe, Phys. Rev. Lett.
{\bf 92}, 020406 (2004).

\bibitem{KetterleMolecules} K. Xu, T. Mukaiyama, J.R. Abo-Shaeer, J.K. Chin, D.E.
Miller, and W. Ketterle, Phys. Rev. Lett. {\bf 91}, 210402 (2003).

\bibitem{Efimov} V.N. Efimov, Sov. J. Nucl. Phys. {\bf 12}, 589 (1971); Nucl. Phys. A
{\bf 210}, 157 (1973).

\bibitem{Recomb} E. Nielsen and J.H. Macek, Phys. Rev. Lett. {\bf 83}, 1566
(1999); B.D. Esry, C.H. Greene, and J.P. Burke, {\it ibid.} {\bf 83}, 1751 (1999).

\bibitem{BraatenRecomb} P.F. Bedaque, E. Braaten, and H.-W. Hammer, Phys. Rev. Lett.
{\bf 85}, 908 (2000).

\bibitem{BraatenAtomDimer} P.F. Bedaque, H.-W. Hammer, and U. van Kolck, Nucl. Phys.
A {\bf 646}, 444 (1999); Phys. Rev. Lett. {\bf 82}, 463 (1999).

\bibitem{EfimovAtomDimer} V.N. Efimov, Sov. J. Nucl. Phys. {\bf 29}, 546 (1979).

\bibitem{VerhaarFeshbach} A.J. Moerdijk, B.J. Verhaar, and A. Axelsson, Phys. Rev. A
{\bf 51}, 4852 (1995).

\bibitem{footnote1} The background scattering length $a_{bg}\sim R_e$ is much smaller
than the resonant part if $|E-E_{res}|\ll\gamma\sqrt{D}$.

\bibitem{LLQ} L.D. Landau and E.M. Lifshitz, {\it Quantum Mechanics}, 
(Butterworth-Heinemann, Oxford, 1999).

\bibitem{footnote2} For a resonance with the magnetic width $\Delta_B$, 
$R^*=\hbar^2/(ma_{bg}\Delta_B\partial E_{res}/\partial B)$.



\bibitem{RempeFeshbach} T. Volz, S. D\"urr, S. Ernst, A. Marte, and G. Rempe, Phys.
Rev. A {\bf 68}, 010702(R) (2003).

\bibitem{RemarkOther} There are narrower resonances in alkalis. We restrict
ourselves to those with magnetic widths $\Delta_B>0.1$G.

\bibitem{FedichevResLight} P.O. Fedichev, Yu. Kagan, G.V. Shlyapnikov, and J.T.M.
Walraven, Phys. Rev. Lett. {\bf 77}, 2913 (1996). 

\bibitem{RemarkTwoPhoton} On two-color photoassociation see J. Weiner, V.S. Bagnato,
S. Zilio, and P.S. Julienne, Rev. Mod. Phys. {\bf 71}, 1 (1999). The scattering
amplitude is derived in J.L. Bohn and P.S. Julienne, Phys. Rev.
A {\bf 60}, 414 (1999).

\bibitem{Skorniakov} G.V. Skorniakov and K.A. Ter-Martirosian, Sov. Phys. JETP {\bf
4}, 648 (1957).

\bibitem{DanilovAtomDimer} G.S. Danilov, Sov. Phys. JETP {\bf 13},
349 (1961).

\bibitem{RemarkFiniteValue} Generally, $f\propto 1+d/r$, but in the absence of a
three-body resonance $d\sim R_e$ and the second term is small.

\end{references}
\end{document}